%% file: main.tex
\def \papermode{0}

\if 0\papermode
\documentclass[aps,pra,reprint,nofootinbib]{revtex4-1}
\else
\documentclass[9pt,twocolumn,twoside]{pnas-new}
\templatetype{pnasresearcharticle}
\fi

\usepackage[utf8]{inputenc}
\usepackage{amsmath,amssymb,bbm,color,lmodern,graphicx}
\newcommand{\stirling}{\genfrac{\{}{\}}{0pt}{}}
\graphicspath{{./images/}}

\if 0\papermode
\begin{document}
\fi

\title{Cooperation risk and Nash equilibrium: quantitative description for realistic players}

\if 0\papermode
\author{G. M. Nakamura}
\email{gmnakamura@usp.br}
\affiliation{Faculadade de Filosofia, Ciências e Letras de Ribeirão
  Preto (FFCLRP) \\
Universidade de S\~{a}o Paulo (USP), 14040-901 Ribeir\~{a}o Preto,
Brazil}
\author{G. S. Contesini}
\email{gcontesini@usp.br}

\affiliation{Faculadade de Filosofia, Ciências e Letras de Ribeirão
  Preto (FFCLRP) \\
Universidade de S\~{a}o Paulo (USP), 14040-901 Ribeir\~{a}o Preto,
Brazil}
\author{A. S. Martinez}
 \email{asmartinez@usp.br}
\affiliation{Faculadade de Filosofia, Ciências e Letras de Ribeirão
  Preto (FFCLRP) \\
Universidade de S\~{a}o Paulo (USP), 14040-901 Ribeir\~{a}o Preto,
Brazil}
\altaffiliation{Instituto Nacional de Ci\^{e}ncia e Tecnologia -
  Sistemas Complexos (INCT-SC)}
\else
\author[a,b]{G. M. Nakamura}
\author[a]{G. S. Contesini}
\author[a,b,1]{A. S. Martinez}
\affil[a]{Faculadade de Filosofia, Ciências e Letras de Ribeirão
  Preto (FFCLRP) \\
Universidade de S\~{a}o Paulo (USP), 14040-901 Ribeir\~{a}o Preto,
Brazil}
\affil[b]{Instituto Nacional de Ci\^{e}ncia e Tecnologia -
  Sistemas Complexos (INCT-SC)}

\leadauthor{Nakamura}
\significancestatement{ %
Realistic description of human behavior in competitive-cooperative
environments remains a significant problem in game theory. Potential
games partially solve this issue by allowing the selection of
sub-optimal strategies by players, reproducing emotional or irrational
responses to decisions. However, potential games sometimes fail to
address the inherent risks associated with cooperation faced by
players. We investigate and solve this issue by introducing a
quantitative measure of cooperation risk, assessing its impact on the
Public Goods game with punishments.}

\authorcontributions{ASM designed research; GMN and GSC performed
  research; GMN wrote the paper; All authors reviewed the manuscript.}
\authordeclaration{The authors declare no conflict of interest.}
\correspondingauthor{\textsuperscript{1}To whom correspondence should
  be addressed. E-mail: asmartinez@usp.br}
\fi

\begin{abstract}
  The emergence of cooperation figures among the main goal of game
  theory in competitive-cooperative environments. Potential games have
  long been hinted as viable alternatives to study realistic player
  behavior. Here, we expand the potential games approach by taking
  into account the inherent risks of cooperation. We show the Public
  Goods game reduce to a Hamiltonian with one-body operators, with the
  correct Nash Equilibrium as the ground state. The inclusion of
  punishments to the Public Goods game reduces the cooperation risk,
  creating two-body interaction with a rich phase diagram, where phase
  transitions segregates the cooperative from competitive regimes.
\end{abstract}

\if 0\papermode
\maketitle

\else

\dates{This manuscript was compiled on \today}
\begin{document}
\verticaladjustment{-2pt}
\maketitle
\thispagestyle{firststyle}
\ifthenelse{\boolean{shortarticle}}{\ifthenelse{\boolean{singlecolumn}}{\abscontentformatted}{\abscontent}}{}
\fi

\input{intro}
\input{public_single_version2}

\input{punishment_single}
\input{conclusion}


\end{document}

%% file: intro.tex
\if 0\papermode
\section{Introduction}
\label{sec:intro}
\fi



Since its initial conception,  Nash equilibrium (NE) has been an
iconic aspect in Game Theory \cite{nashPNAS1950}. It occurs in a game
whenever a player cannot improve her own outcome by changing her
current strategy, under the assumption the remaning players mantain their
respective strategies. This must occur for all the players. Within a
limited set of rules, NE allows for a consistent analysis of
competitive-cooperative scenarios, taking into account the weight of
individual rather than collective performance. The sharp contrast
between NE and cooperative solutions has been the central point in
several studies \cite{axelrodScience1981}. Indeed, it has been shown that
cooperation may emerge as a result of spatial inhomogeneity between
players, modelled with network theory, or a consequence of
more general rules
\cite{szaboPhysRep2007,hauertPNAS2009,szaboPhysRevE2015,szaboPhysRevE2014}.

Players' behavior figures among the relevant aspects governing the game
outcomes. NE assumes players always adopt the best strategy
available to them, reflecting an extensive amount of rational
thinking. Mixed strategies have been sucessful to express players
interactions throught iterated games. However, Game Theory
lacks an exact model.
The formalism of potential games overcome this issue
\cite{blume1993,MONDERER1996124,szaboPhysRep2016} by allowing
players to adopt sub-optimal strategies with a chance governed by a
single parameter $\beta$. The formalism shares a strinking resemblance
with Statistical Physics.
This similarity allows one to borrow tools, interpretations and
results from this discipline and use them into Game Theory analysis 
\cite{szaboPhysRep2016}.


Despite the major advances put forward by potential games, global
rather than individual outcomes play a major role in the quantitative
analysis. While global variables and their minimization
are sensible for physical systems, mostly due to a minimization
principle, the same cannot be said for Game Theory. This
rationale suggests the inherent risks undertaken by players with
cooperative behavior are underestimated. Hence, the resulting
equilibrium may deviates from the NE.

Here, we address the role of NE in the potential game formalism, and
provide an analytical expression for the risks associated with
cooperation. Both are intertwined and necessary to produce a more 
realistic description of outcomes in Game Theory. The paper is
organized as follows. We start employing the Public Goods (PG) game as
our toy model and develop the quantitative description of cooperation
risk. Next, we extend our analysis to include the effect of
punishments. Phase transitions explains the conditions necessary
to create cooperation among players.

%% file: public_single_version2.tex
\if 0\papermode
\section{Cooperation risk}
\else
\section*{Cooperation risk}
\fi

\label{sec:public}

\begin{figure}[htb]
  \if 1\papermode
  \centering
  \fi
  \includegraphics[width=0.4\columnwidth]{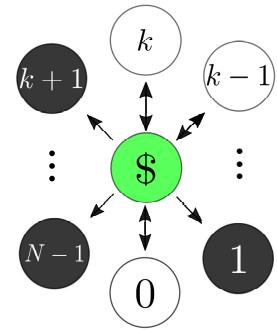}
  \caption{\label{fig:publicplayers} Public goods (PG)
    game with $N$ players. Each player is assigned an unique
    identifier $k=0,1,\ldots,N-1$. Collaborators (white) increase the
    size of public resource by bearing the entry cost $c$ and then
    receive their shares in return (bi-directional arrow). Defector
    (black) keep their assets but still receive share (one-directional
    arrow). }   
\end{figure}

In the Public Goods (PG) games
\cite{hauertPNAS2009,nowakPNAS2009,percSciRep2013}, players may
forfeit the cost $c$ from their own assets to a public resource
(collaborator), or keep $c$ (defector). The public resource is then
increased by the factor $b/c$ and, afterward, redistributed equally
among all players (see Fig.~\ref{fig:publicplayers}). Since players
still receive their share regardless of their own contributions,
defectors lower the returns of collaborators while increasing their
own earnings, thus establishing the cooperative-competitive scenario.

Here, $N$ players partake in a single PG game, with a single resource
pool. 
Each player is labeled by a unique identifier $k=0, 1, \ldots, N-1$
and interacts with $N-1$ different players.
Individual player strategies (cooperator,
defector) are mapped into two-level systems in the Dirac's vector
notation, namely, $\lvert 1 \rangle$ or $ \lvert 0 \rangle$. The
combined strategies of $N$ players creates a configuration vector
$\lvert s\rangle = \lvert n_0 n_1\cdots n_{N-1}\rangle$, where either
$n_k = 1$ (cooperator) or vanishes otherwise (defector). Thus, the
number of distinct configurations is $2^N$, with $\lvert 0 0 \cdots
0\rangle$ being the configuration with defectors (see
Ref.~\cite{nakamuraSciRep2017}).

As usual, player payoffs are crucial to the mathematical description
of the game. They are written in terms of payoff matrices or
operators. Payoff matrices are a common and convenient way to express
players' earnings and costs whenever the number of players is small,
usually $N=2$. For instance, see the payoff matrix of the Prisoner's
Dilemma in Ref.~\cite{hauertEcoLett2005} or \cite{pereiraJTB2010}.
However, as $N$ increases, the  matrix representation becomes
prohibitive. Further insights can be obtained using operators, which
are defined by their action over the configuration vectors, remaining
tractable even for large $N$. In what follows we select the
operatorial description to assess the PG with $N$ players, employing
the hat notation to distinguish operators from numbers.

Our main concern is the action of operator $\hat{n}_k$ over an
arbitrary configuration vector, $\hat{n}_k \lvert n_0n_1\cdots
n_k\cdots\rangle = n_k \lvert n_0n_1\cdots n_k\cdots\rangle $, which
extracts the strategy of the $k$-th player from the configuration
vector. The operators $\hat{n}_k$ ($k=0,1,\ldots, N-1$) hold
additional properties, namely, their eigenvalues are $0$ and $1$; and
they are nil-potent $\hat{n}_k^2 = \hat{n}_k$, making them suitable
building blocks to describe earnings from the various strategies
available to players. More explicitly, the operators that evaluate the
earning of the $k$-th player using cooperative strategies and
defective strategies are, respectively,  
\begin{subequations}
\begin{align}
  \hat{\Pi}^{(C)}_k = 
  & \left[ \frac{b}{N}
    \sum_{\ell=0}^{N-1}\hat{n}_{\ell}
    - c\, \hat{n}_{k} 
    \right]\hat{n}_k ,    \label{eq:pi_c}
  \\
  \hat{\Pi}^{(D)}_k =
  &\left[ \frac{b}{N}
    \sum_{\ell=0}^{N-1}\hat{n}_{\ell}
    - c\, \hat{n}_{k} 
    \right](1-\hat{n}_k).
  \label{eq:pi_d}
\end{align}
\end{subequations}
Since there are only two strategies per player, the total payoff
operator regarding player $k$ is
\begin{equation}
  \hat{\varepsilon}_k=\hat{\Pi}^{(C)}_k+\hat{\Pi}^{(D)}_k=
  \frac{b}{N}\sum_{\ell = 0}^{N-1}\hat{n}_{\ell}   - c\, \hat{n}_{k}.
  \label{eq:earnings_pg}
\end{equation}
Notice that unlike Eqs.~(\ref{eq:pi_c}) and (\ref{eq:pi_d}),
Eq.~(\ref{eq:earnings_pg}) lacks products between operators,
$\hat{n}_k\hat{n}_{\ell}$, the so-called two-body operators. Instead,
Eq.~(\ref{eq:earnings_pg}) holds only one-body operators and, thus,
lacks interactions between different players.

In the formalism of potential games, the Hamiltonian $\hat{H}=
-\sum_k\hat{\varepsilon}_k$ dictates the likelihood of each
configuration according to the Boltzmann distribution
\cite{blume1993,szaboPhysRep2016}.  
One of the key elements of the Boltzmann distribution is the partition
function $Z = \textrm{Tr}({\textrm{e}^{-\beta\hat{H}}})$, which
depends on the parameter $\beta$. In Statistical Physics, $\beta$ is
inversely proportional to temperature. In the context of potential
games, $\beta \in \mathbbm{R}^{+}$ serves as a scale that
model the adoption of sub-optimal strategy by players. With $\beta =
0$, players tend to randomly adopt strategies, whereas
$\beta\rightarrow \infty$ means players tend to adopt the optimal
strategy (rational players). More importantly, by introducing
sub-optimal strategies, the potential games formalism  replaces
mixed strategies to describe the player dynamics. We reinforce that
$\beta$ should be a representative  value for a pool of players
much greater than $N$.

However, we argue $\hat{H}$ fails to correctly
describe the system. Consider the simplest case with $N=2$, which is
formally equivalent to a particular instance of the Prisoner's
Dilemma (PD). In this case,
\begin{equation}
\hat{H}^{(\textrm{PD})} =-\Delta \left(\hat{n}_0+\hat{n}_1\right), \quad \Delta
= b-c,
\label{eq:h_ini}
\end{equation}
where $\Delta$ is the net profit assuming both players
cooperates. Accordingly, the partition function is $Z =
(1+\textrm{e}^{\beta\Delta})^2$, producing the average strategy per
player $\langle n \rangle = (1/2)[1+\tanh(\beta \Delta)]$. Note that
$\beta \gg 1$ and $\Delta > 0$ produce $\langle n\rangle = 1$,
\textit{i.e.}, rational players would cooperate regardless of net
profit as long as $\Delta > 0$. This result is incompatible with the
expected Nash Equilibrium (NE) for $N=2$. Therefore, $\hat{H}$
requires further corrections to take into account the inherent risks
associated with cooperation.

Luckily, the NE requirements can also be used to model the
risk. The condition states the NE occurs whenever a  player cannot
improve her own earnings by changing her current strategy, regardless
of the strategies of the remaining players. We also note that the NE
condition implicitly assumes the various player  strategies are
uncorrelated, to accommodate the assumption of independent 
strategy variations. Let $ \langle \varepsilon_0\rangle =  (b/2)[
\langle n_0\rangle+ \langle n_1\rangle] - c \langle n_0 \rangle$ be
the average earnings of player $k=0$ in a single round PG game with 
two players, with 
$\langle n_{0,1}\rangle \in [0,1]$.  The NE condition reads  $\partial
\langle \varepsilon_0\rangle/\partial \langle n_0 \rangle = (b/2)-c $,
so that increasing cooperation incurs into additional costs 
unless $c < b/2$, with an analogous result for the other
player. Thus, the addition of linear operators $\hat{n}_{0,1}$ with
coupling constants $\mu=\mu_0=\mu_1=c - (b/2)$ incorporates the NE
requirements into the desired PG Hamiltonian with $N=2$:
\begin{equation}
\hat{H}^{(\textrm{PD})} =-{(\Delta-\mu)}
\left( 
\hat{n}_0 +\hat{n}_1
\right).
\label{eq:h}
\end{equation}
It is worth noting that even though $\Delta > 0$, the coupling
$\Delta-\mu$ might acquire negative values. Hence, cooperation becomes
a viable strategy for rational players only if the net return $\Delta$
overcomes the inherent cost $\mu$, associated with cooperation.
Therefore, we define $\mu_k$ as the cooperation risk of player $k$,
and $\mu_k \hat{n}_k$ as the cooperation risk operator.

The NE as the ground state of Eq.~(\ref{eq:h}) can be generalized for
arbitrary $N$. From Eq.~(\ref{eq:earnings_pg})
we evaluate the cooperation risk $\mu_k =-{\partial \langle
  \varepsilon_k \rangle}/{\partial\langle  n_k\rangle} = c -
{b}/{N}$. Due to player translational invariance, $\mu_k\equiv \mu$
and the cooperation risk equals to the net difference between
investment and minimum returns.  Players in PG game aim for increasing
returns while avoiding risks, and are described by the PG Hamiltonian  
\begin{equation}
\hat{H}= -\sum_{k=0}^{N-1}\left(\hat{\varepsilon}_k + \mu_k\hat{n}_{k}
\right)  =-( \Delta-\mu)\sum_{ k=0}^{N-1}\hat{n}_{k},
\label{eq:h_pg}
\end{equation}
with $\mu =c - b/N$.


With Eq.~(\ref{eq:h_pg}) in hands, the partition function $Z_0
=\prod_{k} \left[1+\textrm{e}^{\beta (\Delta-\mu)} \right]$ provides the
average density of cooperators:
\begin{equation}
  \langle n_k\rangle\equiv \bar{n} = \frac{1}{1+ \textrm{e}^{-\beta(\Delta-\mu)}}.
\end{equation}
Hence,  $\Delta  > \mu$ favors cooperation for large values of
$\beta$. Conversely,  $\Delta  < \mu$  inhibits cooperation as
players become aware of risks. Regardless, $\langle
n_jn_k\rangle -\langle n_j\rangle \langle n_k\rangle = 0$, there is
no correlation between players' strategies.

%% file: punishment_single.tex
\if 0\papermode
\section{Cooperation risk in asymmetric games}
\else
\section*{Cooperation risk in asymmetric games}
\fi
\label{sec:punish}

Punishments are socio-economic measures input upon players who disobey
 agreements. Punishments are special because they are asymmetric, only
 affecting a specific subset of players (defectors). They can be
 understood as  adjustment of rules to enforce cooperation. In what
 follows, we explore the PG games with punishment (PGP) to create
 asymmetric payoff operators, introducing correlations among players.

Let us quantify punishment as the  reduction of defectors' earnings by
the factor $0\leqslant \gamma\leqslant 1$. More specifically, the
defector payoff operator in Eq.~(\ref{eq:pi_d}) is modified according
to 
\begin{equation}
  \hat{\Pi}^{(P)}_k=(1-\gamma)\hat{\Pi}^{(D)}_k,
  \label{eq:pi_p}
\end{equation}
which inhibits non-cooperative strategies by decreasing their
effectiveness. The consequence from Eq.~(\ref{eq:pi_p}) appears in the
earning operator regarding the $k$-th player, 
\begin{equation}
  \hat{\varepsilon}_k = -c\,\hat{n}_k+ \frac{b}{N}\left(
    1-\gamma+\gamma\,\hat{n}_k\right)
    \sum_{\ell=0}^{N-1} \hat{n}_{\ell} .
\end{equation}
Due to punishment $\gamma$, $\hat{\varepsilon}_k$ acquires two-body
operators $\hat{n}_{\ell}\hat{n}_k$ with coupling constant
proportional to $\gamma$. Using the property $\hat{n}_k^2 = \hat{n}_k$
 and the guidelines used  in the previous sections, we evaluate the
 cooperation risk  $\mu_k'= \mu_k -\gamma(b/N)\sum\limits_{\ell\neq k}
 \langle n_{\ell}\rangle$ for PGP.
\begin{figure*}[!th]
  \if 1\papermode
  \centering
  \fi
  \includegraphics[width=0.95\columnwidth]{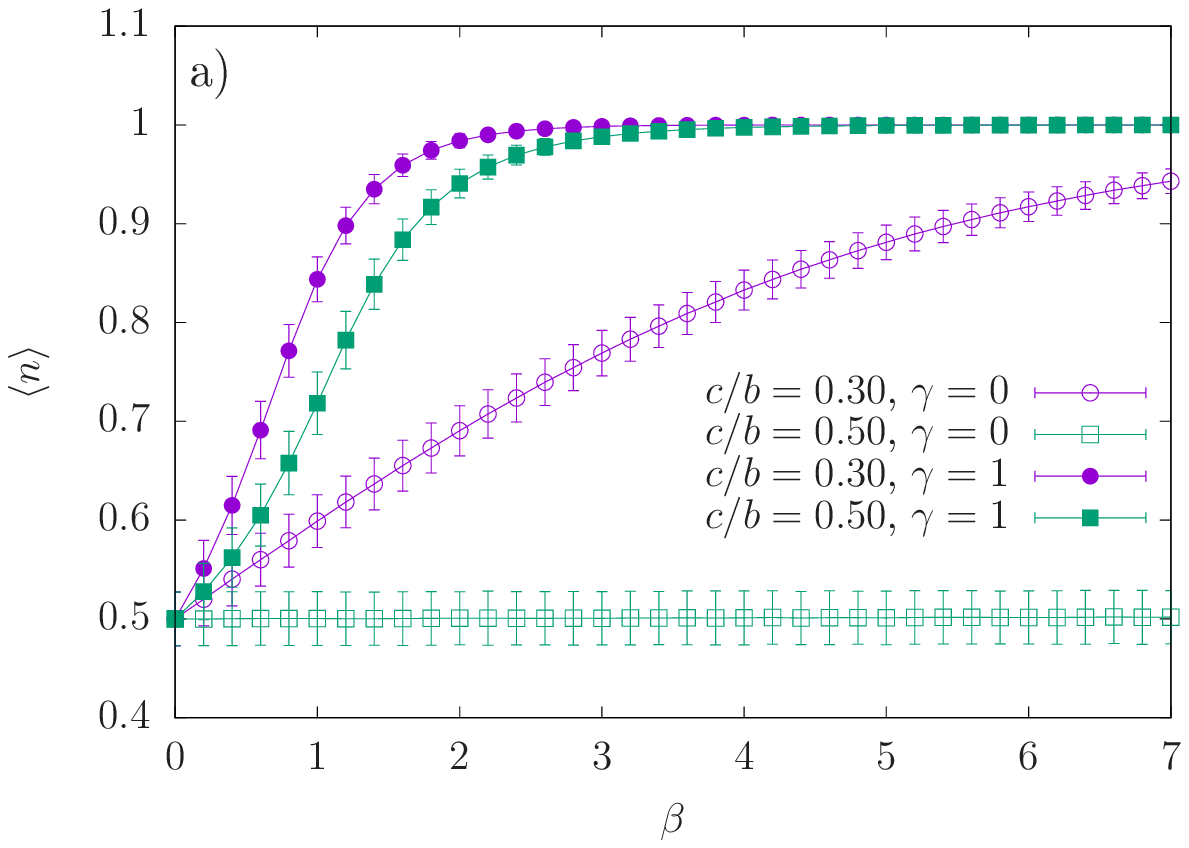}
  \includegraphics[width=0.95\columnwidth]{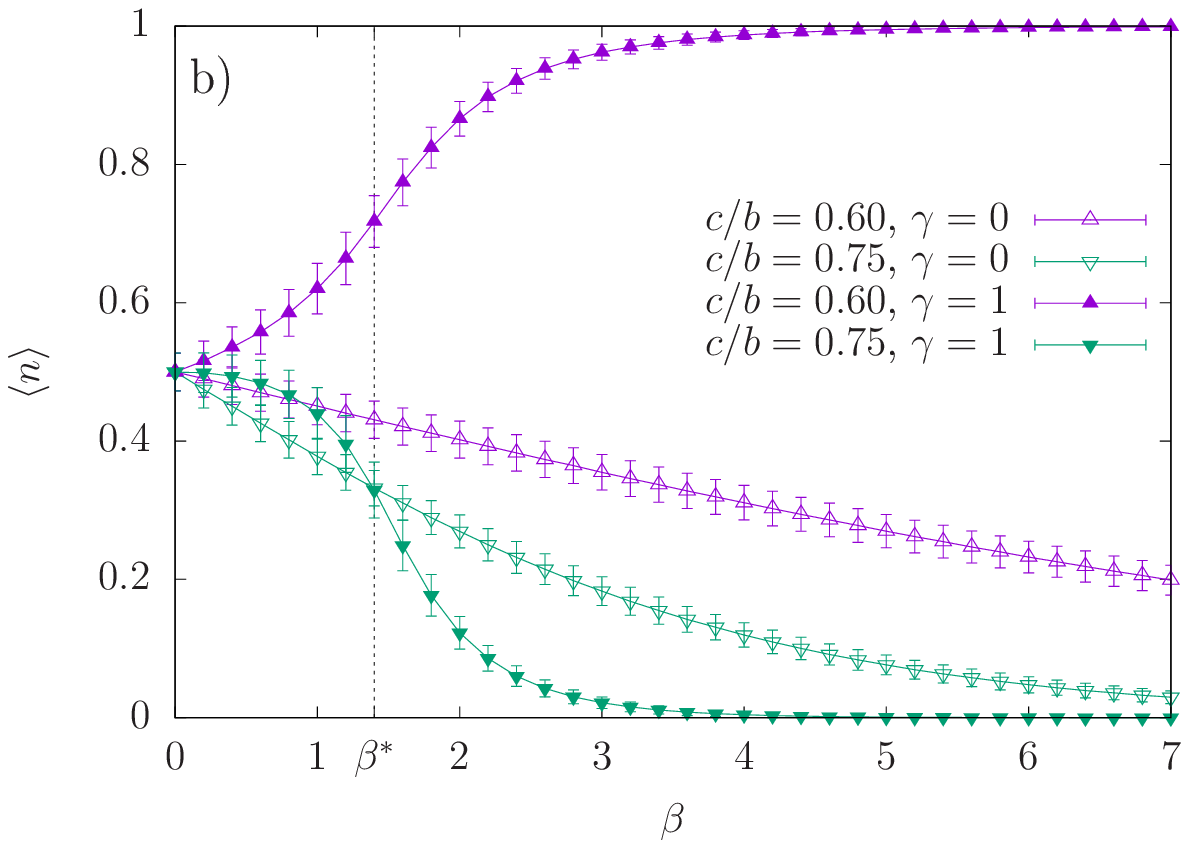}
  \caption{\label{fig:public_punishment} Effects of punishment
    $\gamma$ on $N=1024$ players in a single public goods game.
    a) For $b > 2 
    c$, positive values of $\gamma$ accelerate the adoption
    rate of cooperative strategies. The special case $b=2c$ states
    the equivalence between cooperative and non-cooperative strategies, 
    without punishment. The addition of punishments to the
    game dynamics, however, shifts players towards cooperative
    strategies. b) Punishment retains its efficacy only for a short
    interval of values $2c > b$. In the graphic, for $c/b=0.75$,
    punishment increases the adoption rate of non-collaborative
    strategies, surpassing the case without punishment. The special
    point $\beta^{*}=1.386$ marks the point where the density of
    collaborators with punishment equals its counterpart without
    punishment. Monte Carlo simulations are performed using Metropolis
    algorithm ($10^7$ realizations). Errors bars are estimated
    using integrated correlation time
    \cite{amit2005}.  
  }
\end{figure*}  
Thus, the PGP Hamiltonian with $N$
 players reads 
\begin{equation}
\hat{H}' = - \frac{\gamma b}{N}
  \sum_{\ell,k=0}^{N-1}\hat{n}_{\ell}\hat{n}_k - \sum_{k=0}^{N-1}\left( h_k' - \mu_k'
  \right)\hat{n}_k,\label{eq:pgp_generic}
\end{equation}
where the one-body coupling  $h_k'\equiv h'=\Delta-\gamma b $ differs
by $-\gamma b$ from its counterpart in the PG.

Equation~(\ref{eq:pgp_generic}) supports two remarkable properties.
First, punishment always decreases the risk associated with cooperation:
$\mu'_k - \mu_k = - \gamma (b/N) \sum_{\ell \neq k} \langle n_{\ell}
\rangle \leqslant 0$. Lower risks favor cooperation among players, so
that one may conclude that punishments favor cooperation. However,
punishments also lower the actual value of $h'_k$, which is a primary
component of players' earnings. Payoff decrements $\delta \varepsilon$
due to punishment can be estimated using meanfield approximation:
$\delta \varepsilon \approx -\gamma b \langle n\rangle [1- \langle n
\rangle]$, where $\langle n \rangle \in [0,1]$ describes the mean
cooperation density of players. Therefore, at the same time that
punishment produces a bias towards cooperations, payoffs decrease by
amounts proportional to $\gamma$. Thus, this quantitative result  recovers some
findings first reported in 
Refs.~\cite{nowakPNAS2001,nowakNature2005,nowakScience2009} for
iterated games.

\begin{figure}[htb]
  \if 1\papermode
  \centering
  \fi
  \includegraphics[width=0.95\columnwidth]{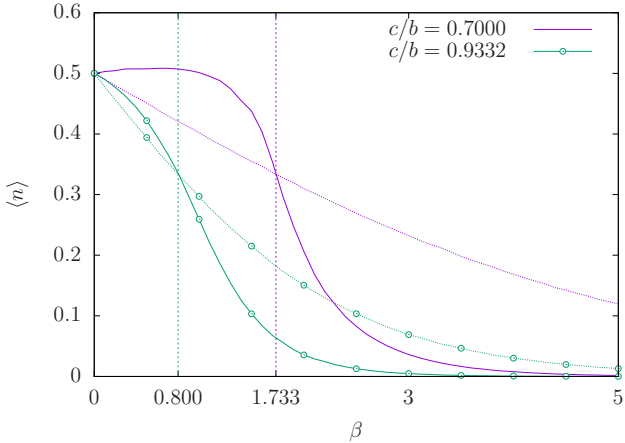}
  \caption{\label{fig:public_punishment2} Equivalence between density
    of collaborators between $\gamma=1$ and $\gamma=0$ for $N=1024$
    players in PGP. The solid lines represent $\langle n\rangle$ with
    $\gamma = 1$, whereas $\gamma=0$ for dashed lines. Vertical dashed
    lines indicate crossing between curves with same cost $c$ but
    different punishment parameters $\gamma$.}
\end{figure}

\begin{figure*}[thb]
  \if 1\papermode
  \centering
  \fi
  \includegraphics[width=0.95\columnwidth]{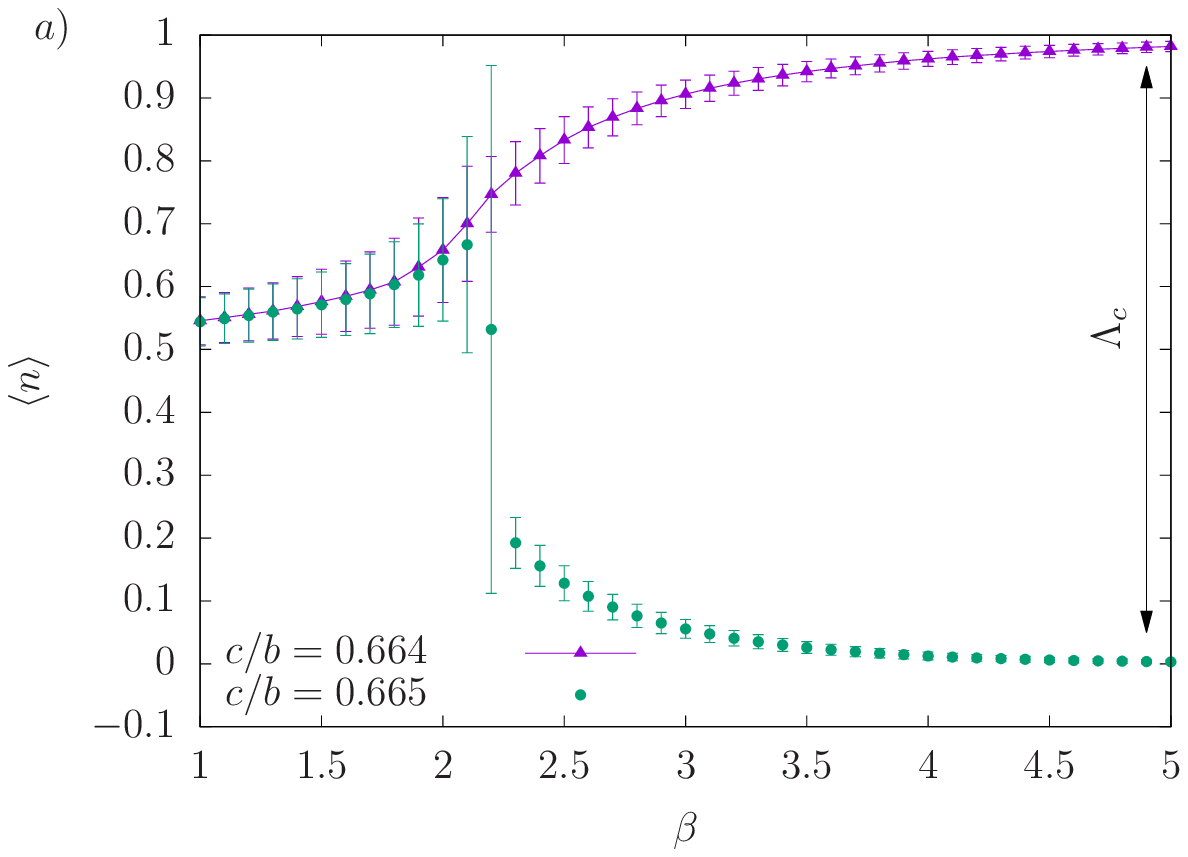}
  \includegraphics[width=0.95\columnwidth]{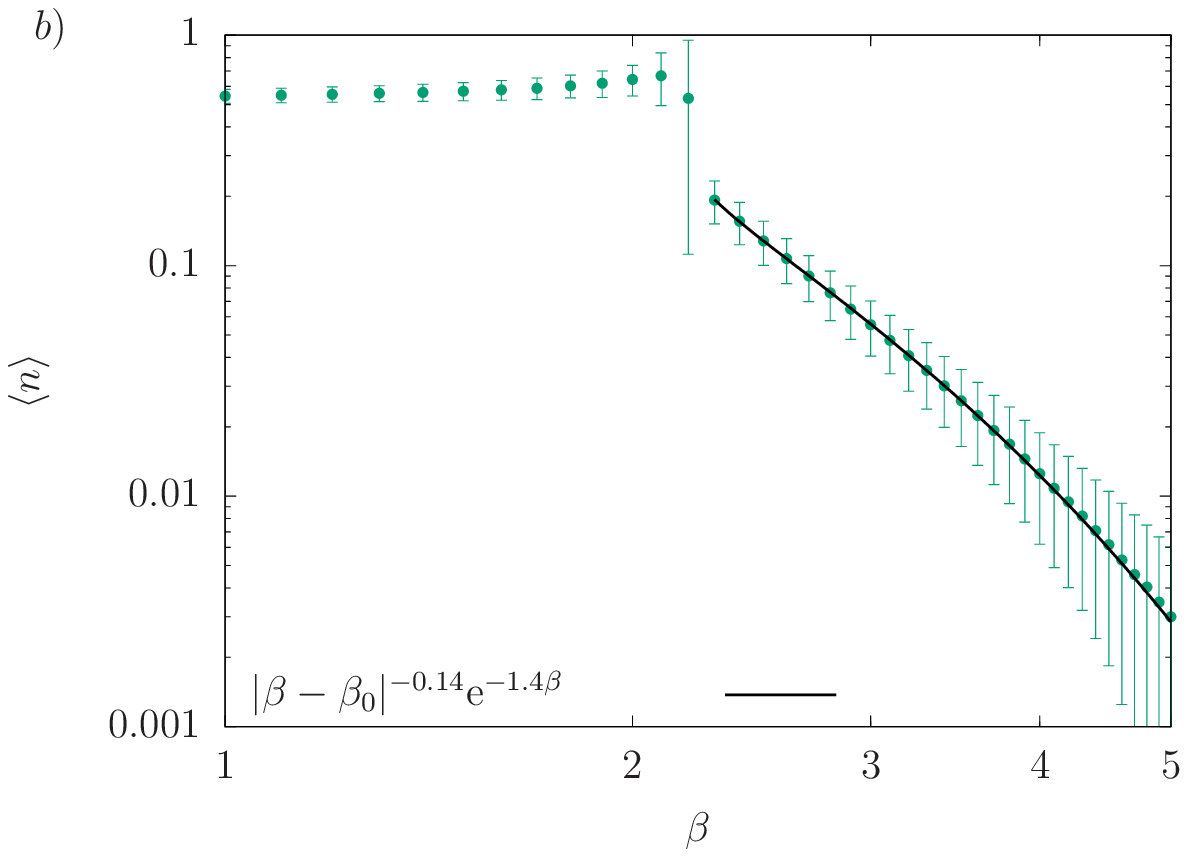}
  \caption{\label{fig:public_punishment3} Phase transition in the
    public goods game with $N=1024$ players, punishment parameter
    $\gamma = 1$ and $b=1$. a) $\langle n  \rangle$ converges
    continously to unity with inverse temperature $\beta$,  for
    $c =0.664$ (triangles). After a small cost increment, player's
    behavior change towards non-cooperation (full circles). The
    gap $\Lambda_c\equiv \Lambda_c(\beta)$ converges to unity for
    large $\beta$. $50\%$ of data omitted for clarity. b)
    $\langle n \rangle$ decays as $|\beta-\beta_0|^{-0.14(0)}\textrm{e}^{-1.40(0)\beta}$ with
    $\beta_0 = 2.175$ (solid line).}   
\end{figure*}

We can learn additional insights about the cooperation risk $\mu_k'$ by
replacing the local average $\langle n_k \rangle$ with the global
average, \textit{i.e.},  $\langle n_k \rangle\approx \langle n\rangle $.
Under the above approximation
\begin{equation}
  \mu_k'\approx \mu_k-\gamma b \langle n \rangle(N-1)/N.
  \label{eq:mu_approx}
\end{equation}
In fact, the approximation becomes exact for translational invariant
systems as players become equivalent to each other: $\mu' = \mu -\gamma b\langle n\rangle (N-1)/N$.
Alternatively, for the sake of practical applications, one may replace
$\langle n \rangle $ in Eq.~(\ref{eq:mu_approx}) by $\bar{n}$, yielding
\begin{equation}
  \mu' = \mu - \left(\frac{N-1}{N}\right)\frac{\gamma
    b}{1+\textrm{e}^{-\beta(\Delta -\mu)}} +o(\gamma^2).
\end{equation}

The second property of Eq.~(\ref{eq:pgp_generic}) concerns two-body
operators $\hat{n}_{\ell}\hat{n}_k$. In general, the overall contribution
attributed to two-body operators depends on the punishment parameter
$\gamma$ and on the local density of cooperators $\langle
n_k\rangle$. Eventually, cooperative strategies become competitive
against the inherent risk associated with cooperation.

To simplify the notation, let the PGP Hamiltonian be written as
$\hat{H} = -\alpha_2\hat{N}^2-\alpha_1\hat{N}$, with
$\hat{N}=\sum_{k}\hat{n}_k$, and couplings $\alpha_2=\gamma b/N$ and
$\alpha_1 = (\Delta -c +b/N) -\gamma  b(1-\bar{n} q/N )$. 
The corresponding partition function reads
\begin{equation}
  Z =Z_0(x) \sum_{k=0}^{\infty}\frac{(\beta\alpha_2)^k}{ k!}\left[
    \frac{1}{Z_0(x)}\frac{\partial^{2k}}{\partial x^{2k}}
     Z_0(x)\right],
  \label{eq:partition_punish}
\end{equation}
where $x=\beta\alpha_1$ and $Z_0(x) = \left( 1+ \textrm{e}^{x}
\right)^{N}$. A more useful formulation for operators $(\partial
/ \partial x)^{2k}$ is obtained after the variable change $u = 1 +
\textrm{e}^{x}$, so that  $(\partial/\partial x)^{2k} =
\sum_{\ell=0}^{2k} \stirling{2k}{\ell}(u-1)^{\ell}(\partial/\partial
u)^{\ell}$. The symbol $\stirling{2k}{\ell}$ refers to the
Stirling  numbers of second kind for $\ell \leqslant 2k$, or $0$
otherwise \cite{abramowitz1964}. 
By applying this expression for $(\partial/\partial x)^{2k}$ to the
term inside the square brackets in Eq.~(\ref{eq:partition_punish}),
one derives the polynomial $\mathcal{P}_{2k}(\xi) =
(\partial/\partial x)^{2k}\ln Z_0$, where
$\xi=(1+\textrm{e}^{-x})^{-1}$. More specifically,
\begin{equation}
  \mathcal{P}_{2k}(\xi)  =   \sum_{\ell=1}^{N} \stirling{2k}{\ell}
  \frac{N!}{(N-\ell)!} \xi^{\ell}.  
\end{equation}

Further algebraic manipulations of Eq.~(\ref{eq:partition_punish}),
with $y \equiv \beta \alpha_2$, produce
\begin{subequations}
\begin{align}
\label{eq:cc0}
Z &= Z_0(x) \left[ 1+
    \mathcal{G}(x,y)\right],\\
\label{eq:cc1}
\mathcal{G} & \equiv
              \sum_{\ell=1}^{N}\frac{N!}{(N-\ell)!}{C_{\ell}(y)}\xi^{\ell},\\
\label{eq:cc2}
C_{\ell}&\equiv \sum_{k=1}^{\infty}\frac{y^k}{k!}
          \stirling{2k}{\ell}.
\end{align}
\end{subequations}
In the special case $\ell = 1$, $C_{1}=
(\textrm{e}^{y}-1)$ grows as the exponential function. 
Under the asymptotic approximation $\stirling{2 k}{\ell} \approx
\ell^{2k}/\ell !$, the functions $C_\ell  \approx 
(\textrm{e}^{y\ell^2}-1)/\ell !$  acquire a much more
tractable form. Note that in both cases, variations in $y$ creates the
same behavior $\delta C_{\ell}=C_{\ell}+ (\textrm{e}^{y
  \ell^2}/\ell!)\delta y+o(\delta y^2)$.

Now, we turn our attention to the average density of collaborators
$\langle n \rangle = (1/N\beta)(\partial /\partial \alpha_1)\ln Z$.
Numerical results from Monte Carlo simulations using Metropolis
algorithm are shown in Fig.~\ref{fig:public_punishment}. The PGP
differs from PG a) by accelerating player compromise rate either \textit{via}
collaboration or defection, depending on the values of $\gamma$ and
$c/b$, for increasing player optimal strategy adoption $\beta$; b)
punishment $\gamma$ promotes cooperation among players for scenarios
that would be otherwise dominated by non-cooperative behavior. From
the analytical point of view, $\langle n \rangle = \bar{n}$ for
vanishing $\gamma$, by construction. In addition, the density $\langle
n \rangle$ satisfies the following inequality:
\begin{equation}
  \langle n \rangle =
  \xi +   \frac{1}{N}\frac{1-\xi}{ 1+\mathcal{G}} \sum_{\ell=1}^{N}\frac{\ell N!
    C_{\ell} \xi^{\ell}}{(N-\ell)!}  \leqslant
  \frac{\mathcal{G}}{1+\mathcal{G}}\left( 1+\frac{\xi}{\mathcal{G}}
  \right).
  \label{eq:density_ine}
\end{equation}
Thus, the density of collaborators meets an upper bound which depends
on $\xi$ and the function $\mathcal{G}$.

Consider the regime of high rationality and low returns,
\textit{i.e.}, $\beta\gg 1$ and vanishing $\xi$. According to 
Eq.~(\ref{eq:density_ine}), there exist three possible outcomes for
$\langle n \rangle$, namely, $\langle n \rangle \rightarrow 0$, if
$\mathcal{G}\rightarrow 0$; $\langle n \rangle \rightarrow 1$, if
$\mathcal{G}\gg 1$; and $\langle n \rangle $ converges to a finite
number in the interval $[0,1]$ if $\mathcal{G}$ converges to a finite
value. These conditions are evident if we consider the largest term in
Eq.~(\ref{eq:cc1}), \textit{i.e.}, $\mathcal{G}\propto
\exp[\beta\bar{\ell}(\gamma\bar{\ell}/N + \alpha_1) ]$. Since
$\alpha_1$ can take negative (positive) values then $\mathcal{G}$ may
vanish (diverge) for large values of $\beta\bar{\ell}$.
Therefore, a non-trivial relationship between $\langle n \rangle$ and
$\gamma$  must emerge under the assumption that the thermodynamic limit
exists for PGP. Indeed, the largest contribution in
Eq.~(\ref{eq:partition_punish}) provides the desired expression:  
\begin{equation}
  2\beta\alpha_2 N\langle n \rangle +\beta \alpha_1 = \Psi( N\langle n
  \rangle+1) - \Psi(N-N\langle n \rangle+1),
  \label{eq:density_condition}
\end{equation}
where $\Psi(z)$ is the Digamma function. Turns out that for fixed
$\gamma$, there exists a cost threshold $c'\equiv c'(\gamma)$ below
which punishment drives cooperative behavior. However, $c > c'$
accelerates the defection rate, leading to the crossing between the
curves $\langle n \rangle $ and $\bar{n}$. In fact, the crossing
occurs at the inverse temperature
\begin{equation}
  \beta^{*}=\frac{\ln 2}{2c-b},
  \label{eq:density_meet}
\end{equation}
obtained from Eq.~(\ref{eq:density_condition}), as shown in
Fig.~\ref{fig:public_punishment2}.

\begin{figure}
  \if 1\papermode
  \centering
  \fi  
  \includegraphics[width=0.95\columnwidth]{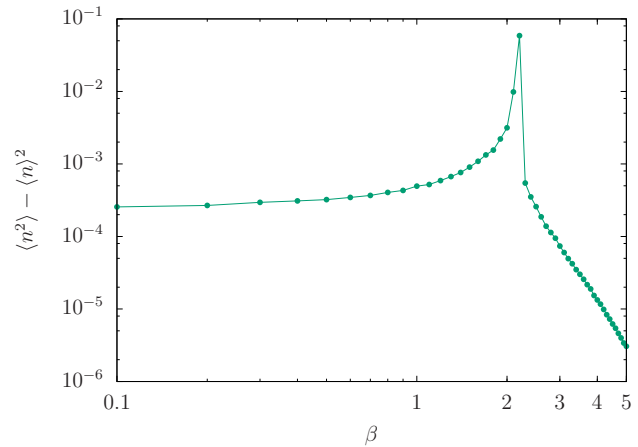}
  \caption{\label{fig:public_punishment4} Variance of cooperation
    density in log-log scale. The maximum occurs around 
    $\beta = 2.2$ for $N=1024$, $c=0.665$, $b=1$ and $\gamma=1$,
    recreating     the shape of $\lambda$ letter, the hallmark
    of $\lambda$-transitions. Error bars omitted for clarity.} 
\end{figure}

Finally, there is the question concerning the value
$c'$. Fig.~\ref{fig:public_punishment3} depicts the behavior of $\langle
n \rangle$ for $c_0=0.664 b$ and $c_1=0.665 b$, with $N=2^{10}$ and
punishment parameter $\gamma=1$. In the first case, punishment tilts
the tendency of players toward cooperation. More importantly, the
cooperator density $\langle n \rangle$ increases monotonically and continuously
for increasing values of $\beta$. In the other case, the new cost
suffers a small increment over the previous cost, $c_1= c_0+\delta c$
with $\delta c = 10^{-3}$. However, the behavior of $\langle n
\rangle$ changes rapidly after $\beta > 2.08$. In fact, numerical
data in Fig.~\ref{fig:public_punishment3} suggests $\langle n \rangle$ 
develops a discontinuity $\Lambda_c = 1$, around $c'\approx 
2/3$ and $\gamma=1$, with defection being the preferred strategy for
players that seek only optimal strategies.
The analogy with Thermodynamics suggests the interpretation of $c$ and
$\langle n \rangle$ as the magnetic field $B$ and magnetization density
$m$, respectively, so that $(\partial m/\partial B)\vert_{\beta ,\gamma} =
\Lambda_c$. This evidence is compatible with a first-order phase
transition,
separating a cooperative phase ($c <
c'$) from a non-cooperative phase ($c \geqslant c'$) for rational
players. Also, we point out that $\langle n \rangle$ changes very
rapidly with $\beta$. A careful analysis in semilog and
log-log scale shows $\langle n\rangle$ decays as
$|\beta-\beta_0|^{-\omega_1}\exp{(-\omega_2\beta)}$, with $\beta_0=2.175$,
$\omega_1 = 0.140\pm 0.002 $, and $\omega_2= 1.400\pm 0.005 \approx
10\, \omega_1 $. More 
importantly, the fluctuation $\langle n^2\rangle  - \langle
n\rangle^2$ displays the well-known shape of $\lambda$-transitions in
log-log scale, as Fig.~\ref{fig:public_punishment4} depicts, with a
peak around $\beta \approx 2.2$. The exact nature of the transition
and whether it occurs as single critical points or rather critical
lines is not entirely clear at this point, being well beyond the scope
of this paper.

%% file: conclusion.tex
\if 0\papermode
\section{Conclusion}
\else
\section*{Conclusion}
\fi
\label{sec:conclusion}


In this paper, a quantitative formulation of cooperation risk is
introduced to the analytical machinery of Game Theory. Cooperation
risk operators are one-body interactions and compete against payoff
operators, and ultimately provide the individuality component
required by the NE. Our numerical results show the PGP develops a
first-order phase transition, separating a cooperative phase from a
non-cooperative phase. Another phase transition is hinted along the
rationality  parameter $\beta$ for specific value of $c/b$ and 
$\gamma$. However, the classification of the phase transition and the
whole range of paramaters in which it occurs is still under study. In
closing, our findings lay out the groundwork for the investigation of
more complex games with competitive-cooperative dynamics, while also
taking into account the individual aspect required by the NE. We plan
to expand this study and evaluate the threshold cost $c'$ for
arbitrary $\gamma$.

\if 0\papermode
  \begin{acknowledgments}
\else
  \acknow{%
\fi  
   We are grateful for F. Meloni comments during manuscript
   preparation and subsequent discussions. GMN holds grant CAPES 
   88887.136416/2017-00, ASM acknowledges grants CNPq
   307948/2014-5. GSC thanks CAPES.
\if 0\papermode   
\end{acknowledgments}
\else
}
\showacknow{}
\fi